\documentclass[]{aastex63}

\usepackage{color}

\received{}
\revised{}
\accepted{}
\submitjournal{ApJ}

\shorttitle{Type IIn Supernova Light Curve}
\shortauthors{Tsuna et al.}

\begin{document}

\title{Type IIn Supernova Light Curves Powered by Forward and Reverse Shocks}

\correspondingauthor{Daichi Tsuna}
\email{tsuna@resceu.s.u-tokyo.ac.jp}

\author{Daichi Tsuna}
\affiliation{Research Center for the Early Universe (RESCEU), School of Science, The University of Tokyo, 7-3-1 Hongo, Bunkyo-ku, Tokyo 113-0033, Japan}
\affiliation{Department of Physics, School of Science, The University of Tokyo, 7-3-1 Hongo, Bunkyo-ku, Tokyo 113-0033, Japan}

\author{Kazumi Kashiyama}
\affiliation{Research Center for the Early Universe (RESCEU), School of Science, The University of Tokyo, 7-3-1 Hongo, Bunkyo-ku, Tokyo 113-0033, Japan}
\affiliation{Department of Physics, School of Science, The University of Tokyo, 7-3-1 Hongo, Bunkyo-ku, Tokyo 113-0033, Japan}

\author{Toshikazu Shigeyama}
\affiliation{Research Center for the Early Universe (RESCEU), School of Science, The University of Tokyo, 7-3-1 Hongo, Bunkyo-ku, Tokyo 113-0033, Japan}\affiliation{Department of Astronomy, School of Science, The University of Tokyo, 7-3-1 Hongo, Bunkyo-ku, Tokyo 113-0033, Japan}



\begin{abstract}
We present a bolometric light curve model of Type IIn supernovae powered by supernova ejecta colliding with a circumstellar medium. 
We estimate the conversion efficiency of the ejecta's kinetic energy to radiation at the reverse and forward shocks and find that a large density contrast makes a difference in the efficiency. The emission from the reverse shock can maintain high efficiency for a long time, and becomes important at the late phase of the light curve. We first construct a semi-analytical model that is applicable to the late phase of the light curve when the diffusion time of photons in the shocked region becomes negligible. We further develop radiation transfer simulations that incorporate these physical processes into the light curve. The numerical calculations predict light curves at early phases, which are testable by present and future short-cadence surveys. We compare our model with the bolometric light curve constructed from observations for a type IIn supernova 2005ip. Due to the reduced efficiency at the forward shock, we find from our model that the mass-loss rate of the progenitor star was $\approx 1\times 10^{-2}\ {\rm M_\odot \ yr^{-1}}$ for a wind velocity of $100\ {\rm km \ s^{-1}}$, an order of magnitude higher compared to previous work that used simple assumptions of the efficiency. This highlights the importance of taking these two components into account when extracting the physical parameters from observations. 
\end{abstract}

\keywords{supernovae: general --- stars: winds, outflows}

\section{Introduction} 
Supernovae, many of which occur upon deaths of massive stars, have a large variety that gives rich information on their diverse evolution. The classification of them is done by an observational perspective, such as spectroscopic features and/or the temporal evolution of their luminosity. 

Among the diversity, a fraction of them present narrow ($\mathcal{O}(10)$ -- $\mathcal{O}(1000)\ {\rm km\ s^{-1}}$) hydrogen emission lines in their spectra. This class of supernovae was recognized in the 1990s and was named as Type IIn \citep{Schlegel90,Filippenko97}. Type IIn supernovae comprise about 10 \% of all core-collapse events (e.g. \citealt{Smith11}). As the width of the emission line reflects the velocity of the matter giving rise to this emission, this feature can not be explained by the supernova ejecta, and indicates the existence of a dense circumstellar medium (CSM) expelled from the progenitor star prior to core collapse. The standard scenario is that a collision between the ejecta and CSM converts the kinetic energy of ejecta to thermal energy via shock heating (e.g. \citealt{Grasberg86,Chugai91,Chugai94,Aretxaga99}). Two shocks called the forward and reverse shocks form, and heat the surrounding material as they pass through. The shock-heated region creates copious photons by e.g. free-free emission, which gradually escape from the shocked region and reaches the observer.

If radiation is mainly supplied by free-free emission, the amount of radiation that can be generated would depend on the density of the shock downstream. Thus the conversion efficiency of the kinetic energy to radiation should in general depend on time. More importantly, the efficiency should be different for the forward and reverse shocks, as the density in the downstreams of the two shocks can be generally very different \citep{Chevalier82}.

There are previous works that have done analytical modelling of the light curve of these interaction-powered supernovae \citep{Chatzopoulos12, Moriya13}, which enable easy extraction of the physical parameters from observations. However, a common caveat in these works is that they do not properly take into account the efficiency at the two shocks. Both of these works assume constant efficiency, and do not take into account the evolution of efficiency in the two components. Although a light curve may be well fit by the models, it can lead to systematic bias in the estimation of physical parameters. There are also numerical works studying ejecta-CSM interaction, in the context of explaining observations of some superluminous supernovae (e.g. \citealt{vanMarle10, Moriya11, Dessart15, Vlasis16,Soumagnac19}). However these works also do not properly consider the efficiency of photon generation and/or the diffusion process of these photons.

One of the physical quantities that can be inferred from observations of Type IIn supernovae is the mass-loss rate of the progenitor just before death. Mass loss of Type IIn progenitors has been historically probed through observations of the intensity and width of the H$\alpha$ emission line \citep{Chugai94,Salamanca98}. Recent compilation of observations reveals (albeit with large uncertainties) huge values of mass-loss rates spanning from $10^{-4}$ to $1\ M_\odot\ {\rm yr^{-1}}$ (e.g., \citealt{Fox11, Kiewe12,Taddia13,Moriya14a}). The extreme mass-loss rates are difficult to be achieved by conventional line-driven wind, and various mechanisms are recently proposed, such as binary interactions (e.g. \citealt{Chevalier12}), pulsational pair-instability predicted for very massive progenitors (e.g. \citealt{WBH07}), core mass-loss due to neutrino emission \citep{Moriya14b}, and gravity waves generated in the convective cores (e.g. \citealt{QS12,SQ14}). Compilation of observations of Type IIn supernovae can help constrain these models, but robustly identifying which mechanism(s) is at play would require accurate modelling of the emission.

In this paper, we construct a model for calculating the bolometric light curve of interaction-powered supernovae by taking into account the structure of the shocked region. We utilize self-similar solutions of spherically symmetric hydrodynamic equations that govern the interaction between a homologously expanding ejecta and a stationary CSM, first obtained by \citet{Chevalier82}. For the case of CSM density profile close to stable mass-loss, as is often assumed in the context of Type IIn supernovae, there is a great difference in density between the inner and outer parts of the shocked region. We find that this density contrast significantly affects the efficiency of converting the kinetic energy to radiation and hence affects the light curve. This would not have been captured by previous works having low resolution inside the shocked region and/or having approximations on the radiation conversion efficiency. 

This paper is constructed as follows. In Section \ref{sec:semiana} we present a semi-analytical model of the bolometric light curve which uses the self-similar solution of \citet{Chevalier82}. In Section \ref{sec:2005ip} we compare our model with the observed light curve of a well-studied Type IIn SN 2005ip, and show that the model can consistently explain the light curve. In Section \ref{sec:simulation} we construct a numerical model using one-dimensional radiation transfer simulations, which can take into account the diffusion of photons. The calculations give predictions of the light curve at the early phase around peak, which can be tested by future surveys. We discuss the underlying caveats of our work in Section \ref{sec:discussion}, and conclude in Section \ref{sec:conclusion}.

\section{Semi-Analytical Model}
\label{sec:semiana}
We construct a semi-analytical model of the light curve of interaction-powered spherical supernovae. This can serve as a model to fit bolometric light curves and obtain parameters of the ejecta and CSM. Our model is applicable to the late phase of the light curve when the diffusion time of photons in the un-shocked CSM becomes negligible compared to the dynamical time scale.

\subsection{Ejecta Profile}
We consider  ejecta in the homologous phase \citep{MM99} when the velocity $v$ can be approximated by $v=r/t$ where $r$ is the radial coordinate and $t$ is time since explosion. The progenitor's radius is ignored in this formulation, but this is a good approximation in our calculations that deal with at least a few days after expansion. The density profile of such ejecta is given in the form of a double power-law function of the velocity as
\begin{eqnarray}
\rho (r,t)= \frac{1}{t^3} \cdot \left\{ \begin{array}{ll}
\left(\frac{r}{gt}\right)^{-n} & (r/t > v_t,\ {\rm outer\ ejecta})\\
(v_t/g)^{-n} \left(\frac{r}{tv_t}\right)^{-\delta}  & (r/t < v_t,\ {\rm inner\ ejecta}),
\end{array}\right.
\label{eq:rho_ej}
\end{eqnarray}
where \citep{Moriya13}
\begin{eqnarray}\label{eq:coeff_ej}
g &=& \left\{\frac{1}{4\pi(n-\delta)} \frac{[2(5-\delta)(n-5)E_{\rm ej}]^{(n-3)/2}}{[(3-\delta)(n-3)M_{\rm ej}]^{(n-5)/2}}\right\}^{1/n}\\
v_t &=& \left[\frac{2(5-\delta)(n-5)E_{\rm ej}}{(3-\delta)(n-3)M_{\rm ej}}\right]^{1/2}.\label{eq:v_t}
\end{eqnarray}
Here $M_{\rm ej}$ and $E_{\rm ej}$ are respectively the mass and (kinetic) energy of the supernova ejecta. The exponent $n$ of the outer ejecta is $\approx 12$ for a progenitor star that has a convective, extended envelope (red supergiants), and $\approx 10$ for a progenitor that has a radiative envelope (blue supergiants and Wolf--Rayet stars). The inner ejecta are considered to have a shallower density profile, with $\delta\sim 0$ -- $1$ for convective progenitors depending on the mass of the envelope, and $\delta\sim 1$ for radiative progenitors \citep{CS89,MM99}. We assume $\delta$ to be $1$ throughout this work. The mass and energy occupied by the outer ejecta can then be calculated by integrating equation (\ref{eq:rho_ej}) as
\begin{eqnarray}
M_{\rm out} = \frac{3-\delta}{n-\delta}M_{\rm ej},\ 
E_{\rm out} = \frac{5-\delta}{n-\delta}E_{\rm ej}.
\label{eq:outer_ejecta}
\end{eqnarray}
\subsection{Hydrodynamics}
Throughout this work we use the self-similar hydrodynamical solutions of ejecta-CSM interaction given in \citet{Chevalier82}. Chevalier's solutions give the hydrodynamical quantities inside the shocked region made by the interaction of homologous ejecta of density profile
\begin{eqnarray}
\rho_{\rm ej} = t^{-3}\left(\frac{r}{gt}\right)^{-n}
\end{eqnarray}
and a CSM (assumed to be stationary) of density profile
\begin{eqnarray}
\rho_{\rm CSM} = qr^{-s},
\label{eq:CSMprofile}
\end{eqnarray}
where $g$ and $q$ are constants. For a stationary wind, $s=2$ and  $q={\dot{M}}/{4\pi v_w}$, where $\dot{M}$ is the mass-loss rate and $v_w$ is the wind velocity. The power-law indices $n$ and $s$ have to satisfy the condition $n>5$ and $s<3$ for a self-similar solution to exist \citep{Chevalier82}. The shocked region has two components, the inner and outer shocked regions created by the reverse and forward shocks. These two regions are separated by the contact surface, where the velocity and pressure are continuous while the density has a discontinuity.

The solution assumes that the shocked region is adiabatic. In reality the leaking of radiation from the shocked region can reduce the pressure in the shocked region and affect the hydrodynamics. If we define an `adiabatic index' $\gamma=\left(\partial\ln P/\partial\ln \rho\right)_s$ in the shocked region, the value will change with time, with $\gamma$ initially being globally $\approx 4/3$ due to the large optical depth and photon pressure, and decreasing towards the isothermal value $\gamma=1$ as the shocked region starts to become optically thin. A self-consistent treatment of the time evolution of this quantity would require radiation hydrodynamics simulations that are computationally expensive. We instead approximate the adiabatic index of the fluid to be constant over time and radius, with a value taking somewhere between these two extremes.

Assuming self-similarity, the hydrodynamical equations inside the shocked region are solved for a given adiabatic index $\gamma$. The Rankine--Hugoniot relations at the two shocks give the boundary conditions, and the requirement of continuous velocity and pressure at the contact surface gives a unique solution that connects the two regions.

\begin{figure}
\centering
\includegraphics[width=0.6\linewidth,angle=-90]{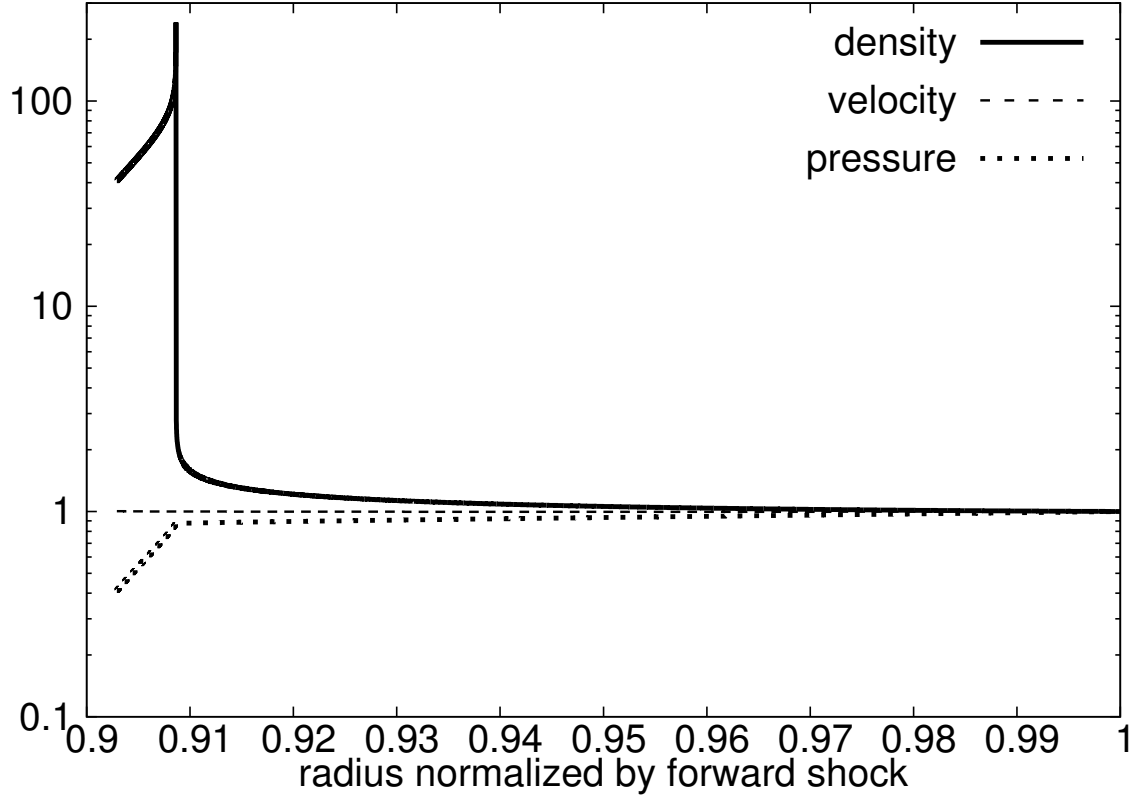}
\caption{Chevalier's self-similar solution for values $n=12, s=2$ and $\gamma = 1.2$. The solid, dashed, and dotted lines show the density, velocity and pressure profiles inside the shocked region. Both axes show values normalized by those at the forward shock.}
\label{fig:Chevalier_n12s2gam65}
\end{figure}

As an example, figure \ref{fig:Chevalier_n12s2gam65} shows the hydrodynamical solutions for the parameter sets $n=12$, $s=2$, and $\gamma=1.2$. The shocked region generally has two components separated by a contact discontinuity, and the density in the shocked ejecta is higher than that in the shocked CSM by a factor of $\sim 40$.

The self-similar solution makes the physical quantities scale as power-law functions of time, whose exponents depend on $n$ and $s$ but not on $\gamma$:
\begin{eqnarray}
r&\propto& t^{(n-3)/(n-s)}, \\
\rho & \propto & t^{-s(n-3)/(n-s)}, \\
v & \propto & t^{(s-3)/(n-s)}, \\
p &\propto& t^{(-sn+5s-6)/(n-s)}.
\end{eqnarray}
The radius and velocity of the contact discontinuity are obtained from dimensional analysis as \citep{Chevalier82}
\begin{eqnarray}
r_{\rm cd} &=& \left[\frac{Ag^n}{q}\right]^{1/(n-s)}t^{(n-3)/(n-s)}, \label{eq:radius_cd} \\
v_{\rm cd} &=& \frac{dr_{\rm cd}}{dt} = \frac{n-3}{n-s} \left[\frac{Ag^n}{q}\right]^{1/(n-s)}t^{-(3-s)/(n-s)},
\label{eq:velocity_cd}
\end{eqnarray}
where $A$ is a constant that depends on $n,s,$ and $\gamma$, and can be numerically obtained (see Table 1 of \citealt{Chevalier82}, for the case of $\gamma=5/3$). Assuming that the CSM is stationary and that both of the shocks are in the strong shock limit, the Rankine--Hugoniot relations at the forward and reverse shocks give
\begin{eqnarray}
v_{\rm fs} &=& \frac{\gamma+1}{2}v_{\rm fs, down},\label{eq:hugoniotv}\\
\rho_{\rm fs, up} &=& \frac{\gamma-1}{\gamma+1} \rho_{\rm fs, down} = qr_{\rm fs}^{-s}, \\
p_{\rm fs, down}  &=& \frac{\gamma-1}{2} \rho_{\rm fs, down}v_{\rm  fs, down}^2,
\end{eqnarray}
and
\begin{eqnarray}
v_{\rm rs} &=&  \frac{n-3}{n-s}  \left[1-\frac{2}{\gamma+1}\frac{3-s}{n-s}\right]^{-1}v_{\rm rs, down} = \frac{n-3}{n-s} v_{\rm ej},\\
\rho_{\rm rs, up} &=& \frac{\gamma-1}{\gamma+1} \rho_{\rm rs, down} = t^{-3}(r_{\rm rs}/gt)^{-n}, \\
p_{\rm rs, down} &=&\frac{2(\gamma-1)}{(\gamma+1)^2} \left(\frac{3-s}{n-s}\right)^2 \left[1-\frac{2}{\gamma+1}\frac{3-s}{n-s}\right]^{-2} \rho_{\rm rs, down}v_{\rm  rs, down}^2.\label{eq:hugoniotp}
\end{eqnarray}
The subscripts 'fs' and 'rs' denote values at the forward and reverse shocks respectively, and the subscripts 'up' and 'down' denote values at the shock upstream and downstream respectively. Newly shocked regions within a time interval $\Delta t$ emit radiation with an energy proportional to the increment in the volume of the shocked region and the radiation energy density at the shock downstream. Thus we obtain the luminosity $L_{\rm fs}$ and $L_{\rm rs}$ of radiation emitted at the two shocks as
\begin{eqnarray}
L_{\rm fs} \Delta t &=& 4\pi r_{\rm fs}^2 (v_{\rm fs}-v_{\rm fs, down})\Delta t \cdot  e_{\rm rad, fs} \nonumber \\
&=& 2\pi (\gamma-1)r_{\rm fs}^2v_{\rm fs, down} e_{\rm rad, fs}\Delta t,\\
L_{\rm rs} \Delta t &=& 4\pi  r_{\rm rs}^2 \left(v_{\rm rs, down} - v_{\rm rs}\right)\Delta t \cdot e_{\rm rad, rs} \nonumber \\
&=& 4\pi \left\{ \frac{n-3}{n-s}  \left[1-\frac{2}{\gamma+1}\frac{3-s}{n-s}\right]^{-1} -1 \right\}r_{\rm rs}^2 v_{\rm rs, down} e_{\rm rad, rs}\Delta t,
\end{eqnarray}
where $e_{\rm rad}$ is the radiation energy density. Similar to \citet{Moriya13}, we assume a thin shell and approximate the radii and velocities of the two shocks as $r_{\rm fs}\approx r_{\rm rs} \approx r_{\rm cd}$ and $v_{\rm fs, down}\approx v_{\rm rs, down} \approx v_{\rm cd}$. Then the densities at the downstreams of the shocks are
\begin{eqnarray}
\rho_{\rm fs, down} &\approx& \frac{\gamma+1}{\gamma-1} \left(\frac{q^{n}}{A^sg^{ns}}\right)^{\frac{1}{n-s}} t^{-\frac{s(n-3)}{(n-s)}}, \\
\rho_{\rm rs, down} &\approx& \frac{\gamma+1}{\gamma-1}\left(\frac{q^{n}}{A^ng^{ns}}\right)^{\frac{1}{n-s}}t^{-\frac{s(n-3)}{(n-s)}},
\end{eqnarray}
the pressures are
\begin{eqnarray}
p_{\rm fs, down} &\approx& \frac{\gamma+1}{2} \left(\frac{n-3}{n-s}\right)^2\left[\frac{q^{n-2}}{A^{s-2}g^{n(s-2)}}\right]^{\frac{1}{n-s}}  t^{-\frac{ns-5s+6}{(n-s)}},\\
p_{\rm rs, down} &\approx& \frac{2}{\gamma+1} \left[\frac{(n-3)(3-s)}{(n-s)^2}\right]^2\left[1-\frac{2}{\gamma+1}\frac{3-s}{n-s}\right]^{-2}\left[\frac{q^{n-2}}{A^{n-2}g^{n(s-2)}}\right]^{\frac{1}{n-s}} t^{-\frac{ns-5s+6}{(n-s)}},
\end{eqnarray}
and the luminosity at the two shocks are given by
\begin{eqnarray}
L_{\rm fs} &\approx& 2\pi (\gamma-1)r_{\rm cd}^2v_{\rm cd} e_{\rm rad, fs} = \frac{2\pi (\gamma-1)(n-3)}{n-s}\left[\frac{Ag^n}{q}\right]^{\frac{3}{n-s}}t^{\frac{2n+s-9}{n-s}} e_{\rm rad, fs},\\
L_{\rm rs} &\approx& 4\pi \left\{ \frac{n-3}{n-s}  \left[1-\frac{2}{\gamma+1}\frac{3-s}{n-s}\right]^{-1} -1 \right\}r_{\rm cd}^2 v_{\rm cd} e_{\rm rad, rs} \nonumber \\
		&=& \frac{4\pi (n-3)}{n-s}\left\{ \frac{n-3}{n-s}  \left[1-\frac{2}{\gamma+1}\frac{3-s}{n-s}\right]^{-1} -1 \right\}\left[\frac{Ag^n}{q}\right]^{\frac{3}{n-s}}t^{\frac{2n+s-9}{n-s}} e_{\rm rad, rs}.
\end{eqnarray}
We will describe how we estimate the radiation energy densities ($e_{\rm rad, fs}$ and $e_{\rm rad, rs}$) at the shock fronts in the next subsection.
\subsection{Radiation Density at Shock}
\label{sec:semiana_erad}
Radiation is supplied at the shock front, where the gas is shock-heated and its energy suddenly increases. This energy is not directly converted to radiation but is first given to ions and electrons, as the timescales of collisions between these particles are shorter than any other relevant timescales.
We calculate the radiation energy density at each shock $e_{\rm rad}$ as follows. The gas at the downstream of each shock is heated up to a temperature of 
\begin{eqnarray}
T_g &\approx& \frac{\mu m_p}{k_B}\frac{p_{\rm down}}{\rho_{\rm down}},
\label{eq:Tgas}
\end{eqnarray}
where $m_p$ is the proton mass, $k_B$ is the Boltzmann constant and $\mu$ is the mean mass per particle in units of $m_p$. We crudely assume here for simplicity that the gas is entirely composed of ionized hydrogen (i.e. $\mu\approx0.5$), but extending this to arbitrary abundances is straightforward. Then we assume electrons emit radiation through free-free transitions\footnote{There are other proposed radiation emission mechanisms, such as synchrotron emission or inverse Compton emission, when the shock becomes collisionless and particle acceleration can occur (see e.g. \cite{Chevalier03}). We will focus on the free-free emission, and consider these other non-thermal emission in future work.} and estimate the energy density of radiation from the pressure at immediately behind the shock front by approximately taking into account effects of the photon diffusion as follows. The free-free emissivity of an ionized plasma of density $\rho_{\rm down}$ and temperature $T_g$ is \citep{Radipro}
\begin{eqnarray}
\epsilon_{\rm ff} &=&  \left(\frac{2\pi k_B}{3m_e}\right)^{1/2} \frac{2^5\pi e^6 \bar{g}_{\rm B}}{3hm_ec^3} n_e T_g^{1/2} \sum_{i} n_iZ_i^2,
\label{eq:eps_ff}
\end{eqnarray}
where $n_e,n_i$ are the electron and ion densities and $Z_i$ is the charge of each ion. For our simple pure hydrogen case $\epsilon_{\rm ff} = Kp_{\rm down}^{1/2}\rho_{\rm down}^{3/2}$, where
\begin{eqnarray}
K &=& \left(\frac{2\pi \mu m_p}{3m_e}\right)^{1/2} \frac{2^5\pi e^6 \bar{g}_{\rm B}}{3hm_em_p^2c^3} \approx 4.7\times 10^{16}\left(\frac{\mu}{0.5}\right)^{1/2}\left(\frac{\bar{g}_{\rm B}}{1.2}\right)\ {\rm cgs\ units}.
\end{eqnarray}
The constants $m_e, e, h, c, \bar{g}_{\rm B}$ are respectively the electron mass,  the electron charge, the Planck constant, the speed of light, and the Gaunt factor taken to be $1.2$. In the case of multiple ions, the contribution from each element should be added. 

At the early phases when $\epsilon_{\rm ff}$ is large enough, gas and radiation will quickly reach thermal equilibrium and the radiation energy density is approximately $3p_{\rm down}$. However at later phases gas and radiation will not reach equilibrium, and the amount of radiation will be limited by $\epsilon_{\rm ff}$, the dynamical time scale over which the pressure and density in the shocked region change significantly, and the diffusion time scale over which photons can be trapped \citep{Nakar10}. As a result, the energy density of radiation is calculated as
\begin{eqnarray}
e_{\rm rad} &\approx& {\rm min}\left[3p_{\rm down}, \epsilon_{\rm ff} {\rm min}(t, t_{\rm diff}) \right] \approx {\rm min}\left[3p_{\rm down}, \epsilon_{\rm ff}\frac{r_{\rm cd}}{c} \right]
\label{eq:eff}
\end{eqnarray}
where we used the fact that we can assume ${\rm min}(t, t_{\rm diff})=t_{\rm diff}\approx r_{\rm cd}/c$ in the optically thin regime. The first term $p_{\rm down}$ evolves as $p_{\rm down}\propto t^{(-ns+5s-6)/(n-s)}$, whereas the second term evolves with a different power-law $\epsilon_{\rm ff}r_{\rm cd} \propto t^{(-2ns+n+7s-6)/(n-s)}$. Which of these two terms decay faster depends on the values of $n$ and $s$. Since $n>s$ is expected to hold for SNe IIn, the second term will decay faster when
\begin{eqnarray}
ns -n - 2s > 0 \iff s > \frac{n}{n-2}.
\end{eqnarray}
If we assume a stationary wind ($s=2$) and $n>5$ (as required from Chevalier's solution), the second term always decays faster than the first term. To proceed with the calculation, we hereafter assume that the second term decays faster.
\subsection{Luminosity at Optically Thin Limit}
According to the argument on the efficiency of emission in the previous section, the radiation power at each shock follows power law temporal evolution with different exponents before and after the transition time when $3p_{\rm down}=\epsilon_{\rm ff}r_{\rm cd}/c$,
\begin{eqnarray}
t_{\rm fs} &=& \left\{\frac{K(n-s)}{3c(n-3)}\sqrt{\frac{2(\gamma+1)^2}{(\gamma-1)^3}}\left[\frac{q^n}{A^sg^{ns}}\right]^{1/(n-s)} \right\}^{1/(ns-n-2s)}
\label{eq:t_fs}
\\
t_{\rm rs} &=& \left\{\frac{K(n-s)}{3c(n-3)(3-s)}\left(1-\frac{2}{\gamma+1}\frac{3-s}{n-s}\right)\sqrt{\frac{(\gamma+1)^4}{2(\gamma-1)^3}}\left[\frac{q^n}{A^ng^{ns}}\right]^{1/(n-s)}\right\}^{1/(ns-n-2s)},
\label{eq:t_rs}
\end{eqnarray}
as
\begin{eqnarray}
L_{\rm fs} = P_{\rm fs} \times \left\{ \begin{array}{ll} \left(t/t_{\rm fs}\right)^{(-ns+2n+6s-15)/(n-s)} & (t < t_{\rm fs})\\
\left(t/t_{\rm fs}\right)^{(-2ns+3n+8s-15)/(n-s)} & (t > t_{\rm fs}),
\end{array}\right. \\
L_{\rm rs} = P_{\rm rs} \times \left\{ \begin{array}{ll}
\left(t/t_{\rm rs}\right)^{(-ns+2n+6s-15)/(n-s)}  & (t < t_{\rm rs})\\
\left(t/t_{\rm rs}\right)^{(-2ns+3n+8s-15)/(n-s)} & (t > t_{\rm rs}),
\end{array}\right.
\end{eqnarray} 
where the constants $P_{\rm fs}$ and $P_{\rm rs}$ can be expressed as
\begin{eqnarray}
P_{\rm fs} &=& \frac{3\pi (\gamma+1)(\gamma-1)(n-3)^3}{(n-s)^3}\left[\frac{A^{5-s}g^{n(5-s)}}{q^{5-n}}\right]^{\frac{1}{n-s}} t_{\rm fs}^{\frac{ns-2n-6s+15}{n-s}}, \\
P_{\rm rs} &=& \frac{24\pi (n-3)^3(3-s)^2}{(\gamma-1)(n-s)^3}\left\{ \frac{n-3}{n-s}  \left[1-\frac{2}{\gamma+1}\frac{3-s}{n-s}\right]^{-1} -1 \right\} \left[\frac{A^{5-n}g^{n(5-s)}}{q^{5-n}}\right]^{\frac{1}{n-s}} t_{\rm rs}^{\frac{ns-2n-6s+15}{n-s}}.
\end{eqnarray}
We define the radiation conversion efficiency at the two shocks $\eta$ as the fraction of the internal energy converted to radiation. Thus
\begin{eqnarray}
\eta_{\rm fs} &=& \frac{L_{\rm fs}}{4\pi r_{\rm fs}^2 v_{\rm fs} p_{\rm fs, down}/(\gamma-1)} \\
\eta_{\rm rs} &=& \frac{L_{\rm rs}}{4\pi r_{\rm rs}^2 (v_{\rm ej}-v_{\rm rs}) p_{\rm rs, down}/(\gamma-1)} 
\label{eq:eta}
\end{eqnarray}
and we apply the thin-shell approximation $r_{\rm fs}\approx r_{\rm rs} \approx r_{\rm cd}$ and $v_{\rm fs, down}\approx v_{\rm rs, down} \approx v_{\rm cd}$. The efficiency will be constant if thermal equilibrium is achieved, but will decrease with time when the free-free emission cannot supply sufficient radiation. We note that because the highest achievable radiation energy density is $3p_{\rm down}$ for each shock, the maximum efficiency in this definition is $3(\gamma-1)$, which is less than unity if $\gamma<4/3$.

Since the luminosity $L_{\rm fs}$ ($L_{\rm rs}$) before the epochs $t_{\rm fs}$ ($t_{\rm rs}$) is proportional to the energy flux incident to the shock front, the time dependence of the light curve is identical to that in \citet{Moriya13} who assumed a constant radiation conversion efficiency of $0.1$ only at the forward shock (i.e. $\eta_{\rm fs} = 0.1, \eta_{\rm rs}=0$). At this stage, the luminosity at the reverse shock is fainter than that at the forward shock because of less incident energy flux to the reverse shock. After these epochs, the reduction of the emissivity results in a more rapid fading of the light curve. As $\epsilon_{\rm ff}$ is proportional to the square of the gas density, the reverse shock is much more capable of maintaining high radiation efficiency than the forward shock. This eventually makes the contribution from the reverse shock important.

When the shocked region becomes optically thin, i.e. radiation free-streams, we can obtain the final form of the light curve as $L(t)=L_{\rm fs}(t)+L_{\rm rs}(t)$. There are five model parameters $g, q, n, s$ and $\gamma$, with which one can attempt to fit a given bolometric light curve. The properties of the ejecta ($M_{\rm ej}, E_{\rm ej}$) lie in a single parameter $g$ as seen from equation (\ref{eq:coeff_ej}), and the properties of the CSM are in the parameter $q$. More information from e.g. spectroscopy, early phase of the light curve, or observation of the progenitor is needed to break the degeneracies among the parameters.

We note that the light curve model is applicable only when the shocked region becomes optically thin. Before that we expect photons to diffuse slowly from the CSM and shocked ejecta, which creates a delay and smoothening in the light curve. The lower limit of $t=t_{\rm min}$ where the model is valid will be when the optical depth from the reverse shock to the observer becomes unity, which can be obtained from the equation 
\begin{eqnarray}
\int_{r_{\rm rs}(t)}^{R_{\rm CSM, out}} \kappa(\rho,T)\rho(r,t) dr = 1
\label{eq:t_min}
\end{eqnarray}
where $\kappa$ is the opacity, and $R_{\rm CSM, out}$ is the outer edge of the CSM. 

Our model does not work after the reverse shock reaches the ejecta core, beyond which Chevalier's solution becomes invalid. This is approximately at the time when the radius of the ejecta core $v_t t$ is equal to the radius of the contact discontinuity $r_{\rm cd}$. From equations (\ref{eq:v_t}) and (\ref{eq:radius_cd}), the upper limit $t_{\rm max}$ is obtained as
\begin{eqnarray}
t_{\rm max} &\approx& \left[\frac{2(5-\delta)(n-5)E_{\rm ej}}{(3-\delta)(n-3)M_{\rm ej}}\right]^{-(n-s)/2(3-s)} \left[\frac{Ag^n}{q}\right]^{1/(3-s)},  \nonumber \\
&\propto& M_{\rm ej}^{(5-s)/2(3-s)} E_{\rm ej}^{-1/2} q^{-1/(3-s)} A^{1/(3-s)},
\label{eq:t_max}
\end{eqnarray}
which is larger for heavier and lower energy ejecta and/or lighter CSM as long as $s<3$.

\section{Application of analytical model}
\label{sec:2005ip}
\begin{figure}
\centering
\includegraphics[width=0.6\linewidth,angle=-90]{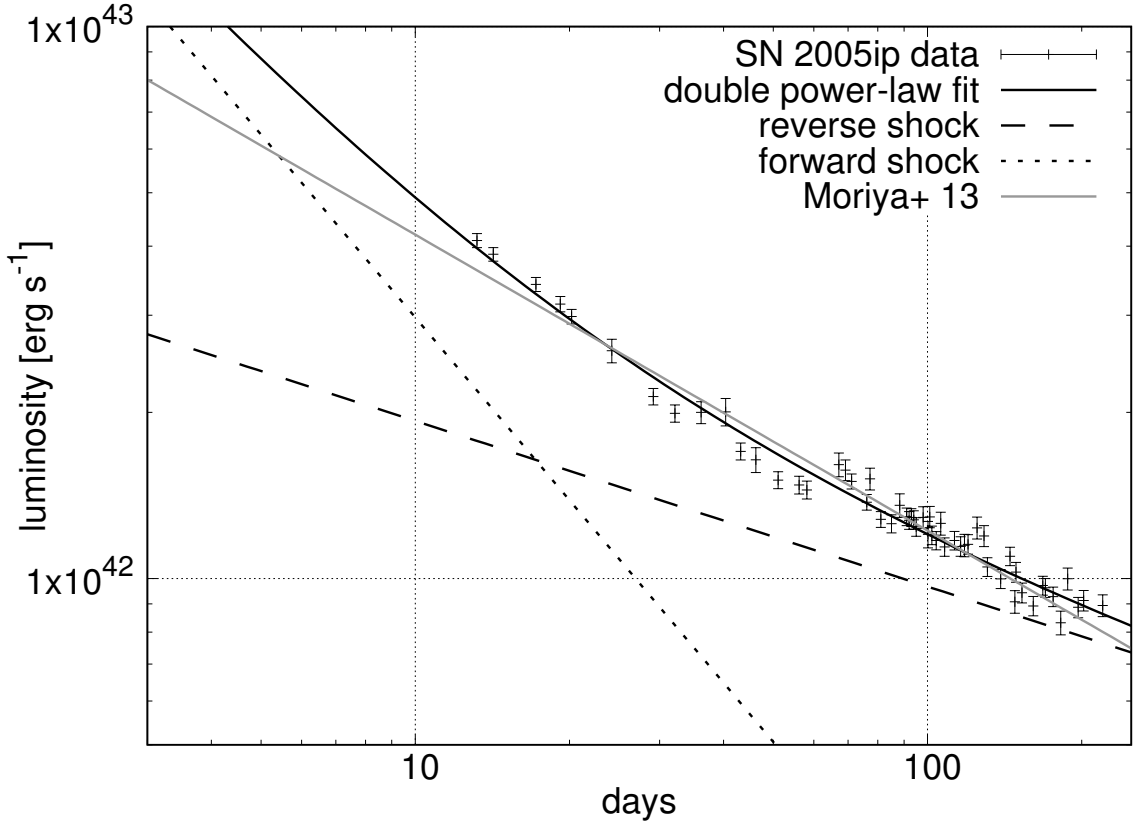}
\caption{Fit of our model of a double-power law bolometric light curve (black line) to observational data for SN 2005ip. A single power-law model with constant efficiency of $0.1$ assumed in \cite{Moriya13} (gray line) is also plotted.}
\label{fig:fitSN2005ip}
\end{figure}

As a demonstration of our semi-analytical model, we consider a double power-law light curve to fit the bolometric light curve of a well-studied Type IIn supernova event SN2005ip \citep{Stritzinger12}. We assume like \citet{Moriya13} that the discovery is 9 days after the explosion, and obtain a least squares fit using data up to 220 days (see Figure \ref{fig:fitSN2005ip}) of the light curve as\footnote{We obtain the reduced chi-squared of the fit as $\approx 2.2$, which is slightly lower than a single power-law fit \citep{Moriya13} of $\approx 3.7$.}
\begin{eqnarray}
L(t) &=& L_{\rm fs}(t) + L_{\rm rs}(t) \nonumber \\
&=& 3.74\times 10^{43}\ {\rm erg\ s^{-1}} \left(\frac{t}{{\rm day}}\right)^{-1.1} + 3.85\times 10^{42}\ {\rm erg\ s^{-1}} \left(\frac{t}{{\rm day}}\right)^{-0.3}.
\label{eq:luminosity_2005ip}
\end{eqnarray}
We have used the density profile with $s=2$ and $n=12$ to obtain the power-law indices, and assumed the former term is the forward shock in the fast-decay phase, whereas the second term is the reverse shock in the slow-decay phase. Using the two coefficients obtained from the fitting, we can obtain the parameters $g,q$ for different values of $\gamma$. 

For example, we numerically find that $A=3.20\times 10^{-2}$ assuming $\gamma=1.2$, and obtain the constants $g,q$ as $g=3.94\times 10^{9}, q=5.13\times 10^{15}$ in cgs units. The parameters $g$ can be translated to $E_{\rm ej}$ by assuming $M_{\rm ej}$, and $q$ can be translated to $\dot{M}$ assuming $v_w$. These values correspond to $E_{\rm ej}=1.28\times 10^{52}$ ergs for a $15\ {M_\odot}$ ejecta, and mass-loss rate of $1.02\times 10^{-2}\ {M_\odot\ {\rm yr}^{-1}}$ for a stable wind of $100\ {\rm km\ s^{-1}}$.

\begin{figure}
\centering
\includegraphics[width=0.6\linewidth,angle=-90]{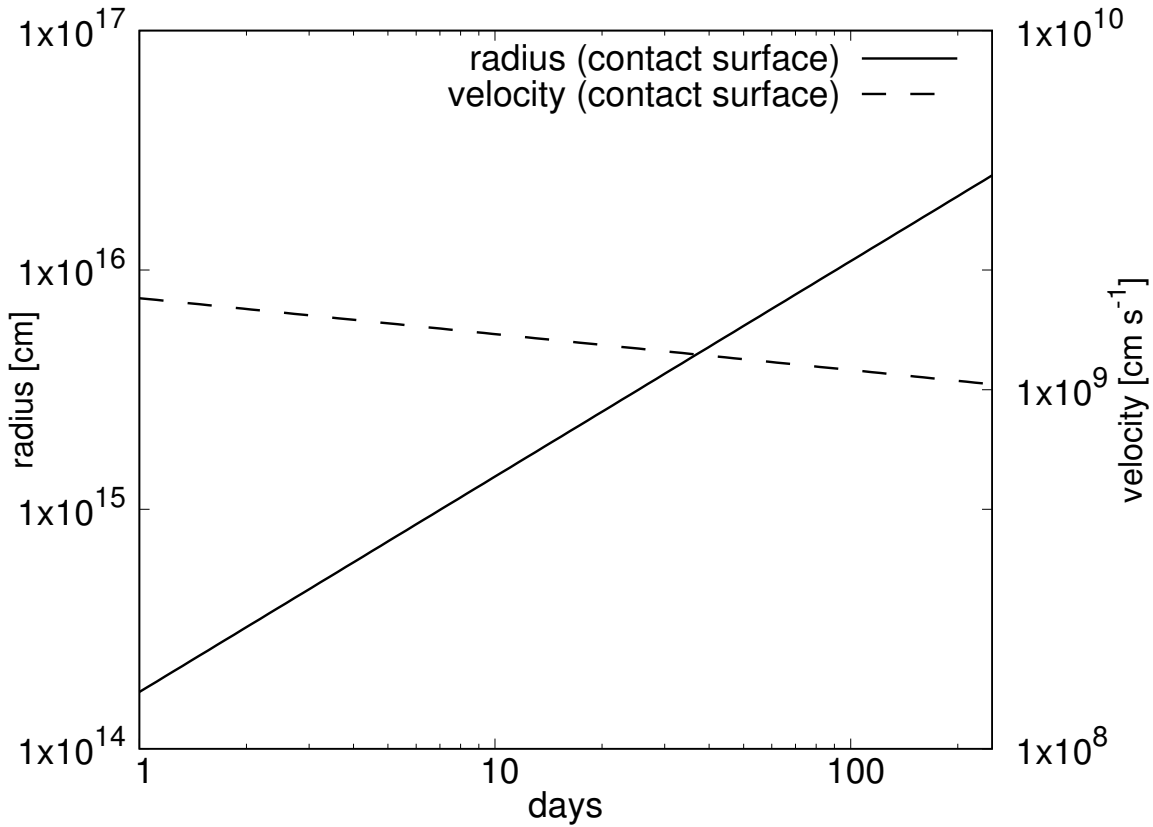}
\caption{Time dependence of radius and velocity of the contact surface, for our model fit to the observational data for SN 2005ip. In our model both of them evolve as a simple power-law, as can be seen from equations (\ref{eq:radius_cd}) and (\ref{eq:velocity_cd}).}
\label{fig:2005ip_radius_velocity}
\end{figure}

In these cases the relevant timescales can be calculated from equations (\ref{eq:t_fs})--(\ref{eq:t_max}) as $t_{\rm fs}\sim0.2$ days, $t_{\rm rs}\sim 300$ days,  and $t_{\rm max}\sim 250$ days. The values of $t_{\rm fs}$ and $t_{\rm rs}$ justify our assumption of a double power-law light curve fit. The observed light curve extending to $\sim 220$ days constrains $t_{\rm max}>220$ days, giving a lower limit to the ejecta mass as $\gtrsim 13\ M_\odot$. This condition with equation (\ref{eq:radius_cd}) also gives a lower limit on the radial extent of the CSM as $\gtrsim 2\times 10^{16}$ cm (see also Figure \ref{fig:2005ip_radius_velocity}), which translates to a CSM mass of $\gtrsim 0.7\ M_\odot$ by integrating equation (\ref{eq:CSMprofile}) and using the fact that the CSM's inner radius is negligible compared to the outer radius.

The lack of observations for the early phases prevents us from constraining $t_{\rm min}$ of this particular supernova. The value of $t_{\rm min}$ is also hard to estimate from our analytical model, due to the difficulty of estimating the value of $\kappa$ inside the shocked region and CSM. The value of $\kappa$ drops drastically when the shocked region cools down by adiabatic and radiative cooling to a temperature of $\sim 6000$ K, where hydrogen starts to recombine. We find from numerical simulations in Section \ref{sec:results} that the temperature inside the shocked region drops to $\sim 6000$ K at $\sim 10$--$20$ days. Thus we conclude that this is the appropriate timescale of $t_{\rm min}$ for this event.

\begin{figure}
\centering
\includegraphics[width=0.6\linewidth,angle=-90]{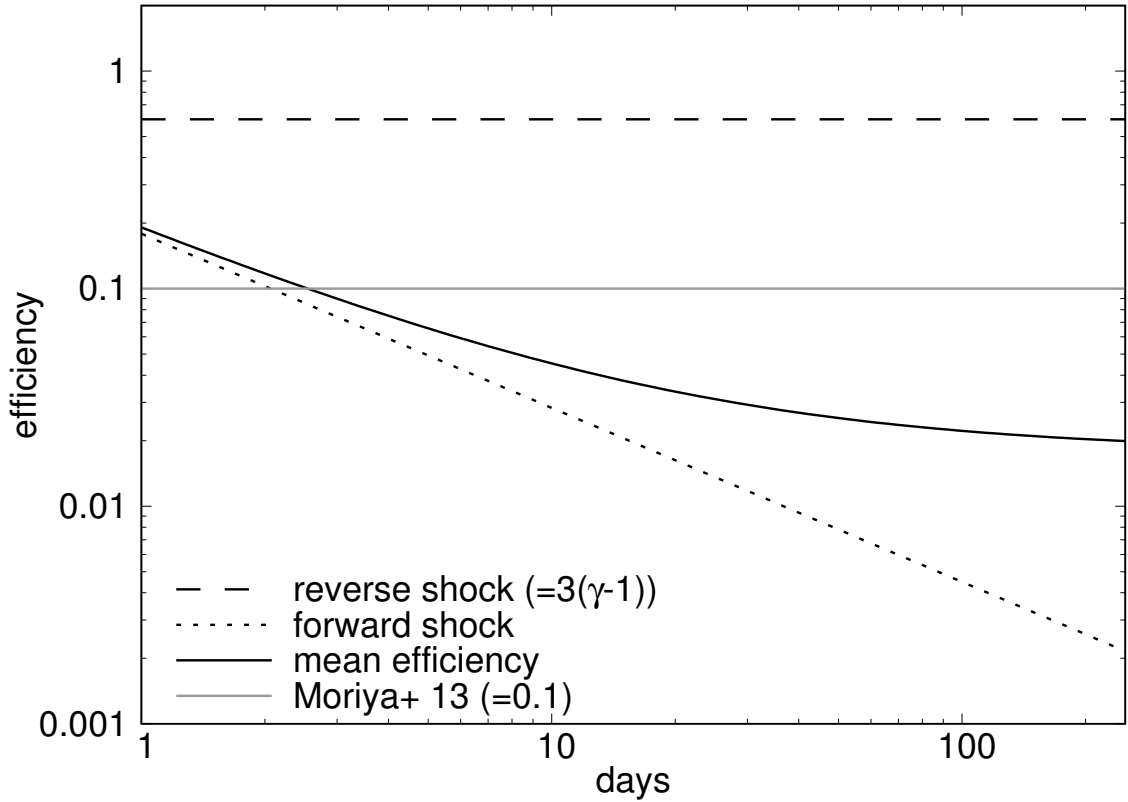}
\caption{Radiation conversion efficiency of the reverse ($=3(\gamma-1)=0.6$) and forward shocks as a function of time for our fitting results for SN 2005ip, and its weighted mean (see main text for the definition). This is to be compared with a power-law model with constant efficiency assumed in \cite{Moriya13} (gray line).}
\label{fig:efficiencySN2005ip}
\end{figure}

Figure \ref{fig:efficiencySN2005ip} shows the evolution of the efficiency in the forward and reverse shocks of our model, and the weighted mean, which is the actual luminosity in equation (\ref{eq:luminosity_2005ip}) divided by the luminosity if the efficiencies in both shocks were unity, i.e.
\begin{eqnarray}
\eta_{\rm mean} = \frac{L_{\rm fs}+L_{\rm rs}}{L_{\rm fs}/\eta_{\rm fs} + L_{\rm rs}/\eta_{\rm rs}}
\end{eqnarray}

The mean efficiency $\eta_{\rm mean}$ decays over time and approaches the value $\sim 0.02$, which is the ratio of kinetic energy dissipated in the reverse and forward shocks. We find that the efficiency is roughly consistent with the standard value of $0.1$ adopted by \citet{Moriya13}.

However, the most important thing is that this efficiency evolves over time, and this should be taken into account in modelling the light curve. In fact, it is noteworthy that the required mass-loss rate we obtained is about an order of magnitude higher than that obtained in \cite{Moriya13}. This is because the forward shock cannot sustain the high efficiency as assumed in their paper (see Figure \ref{fig:efficiencySN2005ip}), and the reverse shock, while possible to maintain a high efficiency, dissipates much less kinetic energy than the forward shock. The difference in the mass-loss rates stress the importance of taking both of these two shocks into account when modelling the light curve.

Our estimate of the mass-loss rate is roughly in line with estimates from other independent observations of this event \citep{Fox10, Fox11, Katsuda14}. An estimate by \cite{Smith09} give a much lower mass-loss rate of $\approx 2\times 10^{-4}\ M_\odot\ {\rm yr^{-1}}$, but we attribute this apparent tension to the much higher radiation conversion efficiency ($=0.5$) assumed in their work.

\section{Simulation}
\label{sec:simulation}
Our analytical model described in the previous sections neglects diffusion of photons in the shocked ejecta and CSM. In this section, we present our method of Lagrangian numerical simulations to calculate the light curve by taking into account this diffusion process. To do this, we numerically solve the radiative transfer equation using a flux limited diffusion approximation with the energy dissipation rates at the shock fronts obtained by following the procedures in Section \ref{sec:semiana_erad}. We give a detailed description of our method in the following subsections.

\subsection{Hydrodynamics}
We obtain the hydrodynamical evolution of the interaction region utilizing Chevalier's self-similar solution. Since the values of the exponents appearing in the solution are restricted as $n>5$ and $s<3$, the solutions cannot be used after the reverse shock enters the inner ejecta where $n=1$. We thus stop the calculations of the light curve at this epoch. Afterwards, it is expected that the luminosity will asymptotically decay with time as the kinetic energy supply from the ejecta decreases \citep{Moriya13}. It is unclear what happens to the light curve at the transition region, where the behaviour of the reverse shock drastically changes. To probe this regime in detail, self-similar solutions cannot be used and a suite of hydrodynamical simulations would be necessary.

\subsection{Radiation Transfer}
Using the hydrodynamical solutions as background, we adopt a simple scheme of radiation transfer and solve the radiation flow within the shocked region and the unshocked CSM. We solve radiation transfer in the computational cells using the formulation of flux-limited diffusion given in \citet{LP81}
\begin{eqnarray}
F = - \frac{\lambda c}{\kappa\rho} \frac{\partial (aT^4)}{\partial r},
\label{eq:LP_F}
\end{eqnarray}
where $a$ is the radiation constant and
\begin{eqnarray}
\lambda &=& \frac{2+|R|}{6+3|R|+|R|^2} \\
R &=&  -\frac{1}{\kappa\rho}\frac{\partial (aT^4)/\partial r}{aT^4}.
\label{eq:LP_R}
\end{eqnarray}
The value of $\lambda$ becomes $1/3$ in the optically thick limit, whereas it approaches $1/|R|$ in the optically thin limit. We assume the flux at the inner computational boundary is zero, and free streaming with $F=acT^4$ at the outer boundary. At each time step, the temperature in each cell is correspondingly updated by adiabatic cooling and radiation transfer using the first law of thermodynamics:
\begin{eqnarray}
\frac{\dot{T}}{T} = \frac{1}{3}\frac{\dot{\rho}}{\rho} - \frac{1}{4aT^4} \cdot \frac{1}{r^2} \frac{\partial}{\partial r} (r^2 F),
\end{eqnarray}
where the dot denotes the Lagrangian time derivative and the time derivative of $\rho$ is given by Chevalier's solution. In this work we neglect contribution from other heating sources, such as radioactive decay of $^{56}$Ni. This assumption is supported by the relatively low $^{56}$Ni mass inferred from observations of Type IIn supernovae (\citealt{Elias-Rosa18} and references therein).

We assume the total (absorption $+$ scattering) opacity $\kappa$ to be the Rosseland mean opacity, and use the values in the OPAL opacity table for the solar abundance \citep{Iglesias96}. We take the radiation temperature as input, which implicitly assumes local thermodynamic equilibrium (LTE) throughout the computational region. Although this becomes a bad approximation as radiation and gas start to decouple, we expect this assumption  not to greatly affect our results on the bolometric light curve as radiation tends to freely stream in this regime for both cases anyway.

We assume electron scattering to be the dominant source of the scattering opacity, and independently obtain the scattering opacity $\kappa_{\rm scat}$ by solving Saha's equations on the ionization of hydrogen and helium at solar abundance. The absorption opacity is then defined as $\kappa_{\rm abs} = \kappa-\kappa_{\rm scat}$. For cells of densities lower than the edge of the table, corresponding to densities of $[\rho/({\rm g\ cm^{-3}})]/(T/10^6{\rm K})^3 < 10^{-8}$, we extrapolate the absorption opacity from the edge of the table with the assumption that the absorption opacity is proportional to the density. For cells of temperatures lower than the edge of the table (corresponding to $T\lesssim 5600$ K), we neglect the contribution from the absorption opacity. For these two exceptional cases the total opacity $\kappa$ is defined as the sum of this revised absorption opacity and the scattering opacity.

\subsection{Radiation Density at Shock}
In our Lagrangian scheme, there will be cells that are initially in the unshocked CSM but eventually swallowed by the forward shock. When the forward shock propagates into a new unshocked CSM cell in front, we "shock" this cell by giving a density $\rho_{\rm down}$ and a pressure $p_{\rm down}$ obtained from the self-similar solution (eqs. \ref{eq:hugoniotv}-\ref{eq:hugoniotp}). The resulting radiation supplied to the shocked region is calculated from the emissivity in equation (\ref{eq:eps_ff}), with gas assumed to be fully ionized and of solar abundance, i.e. $\mu\approx 0.62$. We consider hydrogen and helium for the ions, and assume all of hydrogen and helium are ionized to obtain $n_e$ and $n_i$. Similar to equation (\ref{eq:eff}) we define the (radiation) temperature $T$ of the newly shocked cell with the equation
\begin{eqnarray}
aT^4 = {\rm min}[3p_{\rm down}, \alpha \epsilon_{\rm ff} \cdot {\rm min}(t, t_{\rm diff})],
\label{eq:rad_efficiency}
\end{eqnarray}
where $t_{\rm diff}= {\rm max}(\tau,1)r/c$ is the diffusion time scale, and $\alpha$ is a parameter, assumed to be constant for simplicity, that determines the efficiency of thermalization in the shocked downstream. We measure the optical depth $\tau$ at each shock front. The value of the parameter $\alpha$ should depend on the micro-processes that occur at the shock. Here we adopt $\alpha=1$ for our light curve calculations, and study in Section \ref{sec:alpha_gamma} the dependence of the light curve on its value.

\subsection{Parameter Sets and Results}
\label{sec:results}
In this work we consider six parameter sets to study the dependence of our light curve on ejecta and CSM properties. The parameter sets are summarized in Table \ref{table:IInParameters}. Because the self-similar solution requires constant $\gamma$, we set the hydrodynamical background assuming $\gamma=1.2$, and study the dependence of our results of the light curve on this value in Section \ref{sec:alpha_gamma}.

\begin{table*}
\centering
\begin{tabular}{c|cccccc}
Model & $n$ & $E_{\rm ej}\ ({\rm erg})$ & $\dot{M}/v_w\ ((M_\odot/{\rm yr})/({\rm km/s}))$ & $M_{\rm CSM}\ (M_\odot)$ & $M_{\rm sh, 14}\ (M_\odot)$ & $v_{\rm fs, 14}$ ({\rm km/s}) \\ \hline
Fiducial  & $10$& $10^{51}$ & $3\times 10^{-2}/100$ & $1.9$ & $3.7\times 10^{-2}$& $6.4\times 10^3$\\ 
High-CSM & $10$& $10^{51}$ & $6\times 10^{-2}/100$ & $3.8$ & $7.4\times 10^{-2}$ & $5.8\times 10^3$\\
Low-CSM & $10$& $10^{51}$ & $1\times 10^{-2}/100$ & $0.6$ & $1.2\times 10^{-2}$ & $7.5\times 10^3$\\
High-$E$& $10$& $3\times 10^{51}$ & $3\times 10^{-2}/100$ & $1.9$ & $3.7\times 10^{-2}$ & $1.1\times 10^4$\\
Low-$E$ & $10$& $3\times 10^{50}$ & $3\times 10^{-2}/100$ & $1.9$ & $3.7\times 10^{-2}$ & $3.5\times 10^3$\\
n12 & $12$ & $10^{51}$ & $3\times 10^{-2}/100$ & $1.9$ &  $4.5\times 10^{-2}$ & $5.6\times 10^3$
\end{tabular}
\caption{Six sets of model parameters tested in our simulation. The first four columns represent: power-law index of the ejecta density, energy of the ejecta, ratio of the mass-loss rate and the wind velocity, and total CSM mass. The last two columns are respectively the mass of the shocked region and velocity of the forward shock when the contact discontinuity is at $10^{14}$ cm. For all the parameter sets, the ejecta mass is $10\ M_{\odot}$, the CSM power-law index is $2$, the outer edge of the CSM is $2\times 10^{16}$ cm, $\alpha$ is set to $1$, and the adiabatic index is set to $1.2$.} 
\label{table:IInParameters}
\end{table*}

We start our simulations with an initial radius of the contact discontinuity at $10^{14}$ cm. The computational region is set to be the shocked region, and the unshocked CSM which extends out to a radius of $2\times 10^{16}$ cm. Each cell is set to have an equal mass $\Delta m_{\rm in}$ in the inner shocked region, and $\Delta m_{\rm out}$ in the outer shocked region and CSM. The latter is a few times lower than $\Delta m_{\rm in}$. The total number of cells is changed for each model parameter, and grows over time as the reverse shock propagates through the ejecta, but generally stays in the order of 1000 for all cases.

As shown in Table \ref{table:IInParameters}, when the shocked region is at the initial radius $10^{14}$cm it has a mass of the order of $M_{\rm sh, 14}\sim 10^{-2}\ M_\odot$. At this stage, the shocked region is optically thick with an optical depth of
\begin{eqnarray}
\tau &\sim& \frac{\kappa M_{\rm sh}}{4\pi r^2} \sim 200 \left(\frac{\kappa}{0.34\ {\rm cm^2\ g^{-1}}}\right) \left(\frac{M_{\rm sh, 14}}{3.7\times 10^{-2}M_\odot}\right) \left(\frac{r}{10^{14}\ {\rm cm}}\right)^{-2}.
\end{eqnarray}
The photon diffuses through the shocked region at a velocity
\begin{eqnarray}
\frac{c}{\tau} \sim 1.5\times 10^3\ {\rm km\ s^{-1}} \left(\frac{\tau}{200}\right)^{-1} \lesssim v_{\rm sh, 14},
\end{eqnarray}
where $v_{\rm sh,14}$ is the velocity of the forward shock at $r=10^{14}$cm. As the photon diffusion is slower than the forward shock at this stage, we can assume that (i) the gas is adiabatic and that (ii) radiation and gas are coupled. We thus determine the initial radiation temperature $T$ of each cell in the shocked region from the pressure and density in the self-similar solution by
\begin{eqnarray}
p_{\rm Ch} = \frac{a}{3}T^4 + \frac{\rho_{\rm Ch} k_BT}{\mu m_p},
\label{eq:eos_ideal}
\end{eqnarray}
where $p_{\rm Ch}$ and $\rho_{\rm Ch}$ are the pressure and density obtained from Chevalier's self-similar solution and $a$ is the radiation constant. We assume $\mu=0.62$ for all cases. The initial radiation temperature inside the unshocked CSM is uniformly set to be $500$ K, a temperature low enough that the radiation stored in the CSM is negligible compared to that in the shocked region.

\begin{figure}
\centering
\includegraphics[width=0.6\linewidth,angle=-90]{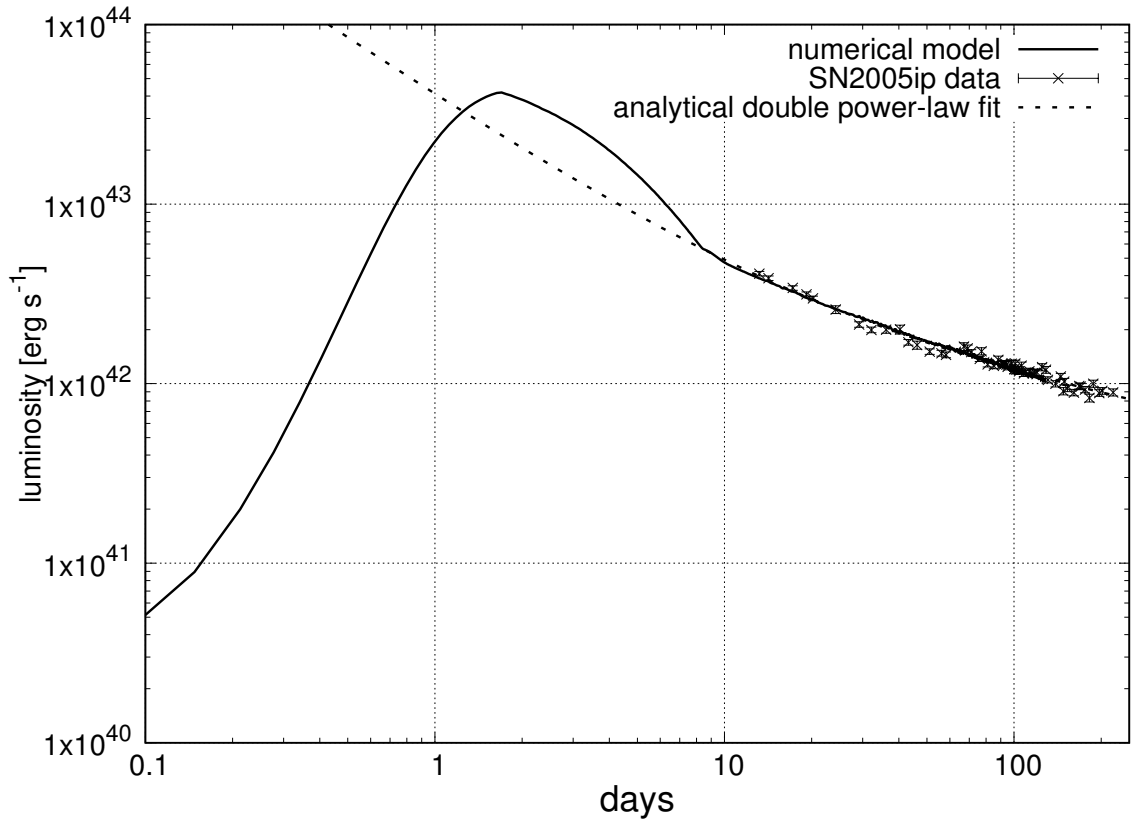}
\caption{Light curve calculated from our numerical model out to $100$ days, plotted with the two-component power-law fit in Figure \ref{fig:fitSN2005ip} with dotted lines.}
\label{fig:SN2005ip_numerical}
\end{figure}

We have first done numerical calculations to reproduce the observed light curve of SN 2005ip, shown in Figure \ref{fig:SN2005ip_numerical}. The calculated light curve matches the observations, and the double power-law fit calculated from the analytical model in Section \ref{sec:2005ip} at late phases. We used the same $n, s, \gamma$ and mass-loss rate with the analytical model, but adopted $g\approx 4.2\times 10^9$, which corresponds to an explosion energy of $\approx 1.5\times 10^{52}$ erg for 15 $M_\odot$ ejecta. This required energy is about 20 per cent higher than what was predicted from the analytical model. This can be qualitatively explained from the fact that we have underestimated the radius of the forward shock $r_{\rm fs}$ by $\sim 10$ per cent in our analytical model, by approximating it as equal to $r_{\rm cd}$. This leads to an overestimation of the density at forward shock downstream ($\propto r_{\rm fs}^{-s}$), which greatly enhances the free-free emissivity ($\propto r_{\rm fs}^{-2s}$). Although the luminosity is also affected by the underestimation of $r_{\rm fs}$, the enhancement of emissivity has an even larger effect.

\begin{figure}
 \centering
\includegraphics[width=0.6\linewidth, angle=-90]{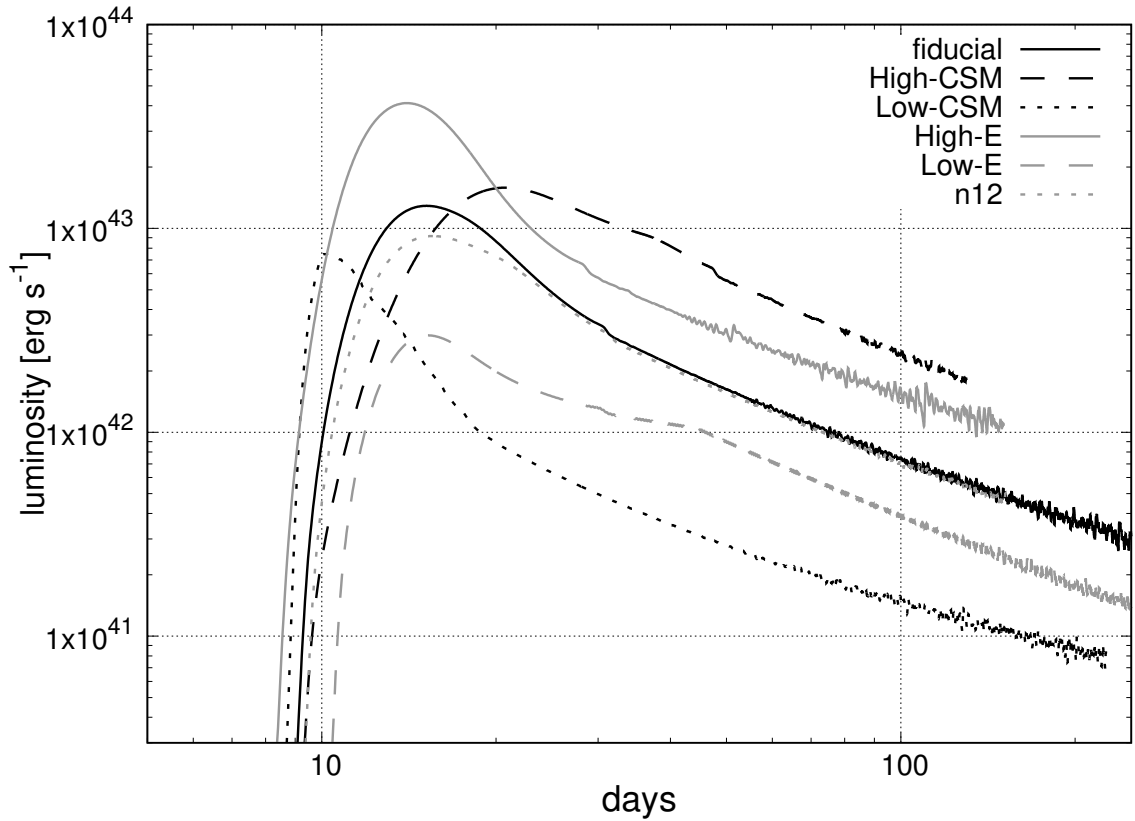}
 \caption{Light curves for the six parameter sets shown in Table \ref{table:IInParameters}. The light curves are terminated either when the forward shock reaches the outer edge of the computational region, or when the reverse shock reaches the inner edge of the outer ejecta.}
\label{fig:lightcurves}
\end{figure}

The light curves for the other parameter sets in Table \ref{table:IInParameters} are shown in Figure \ref{fig:lightcurves}. We see that the light curves generally have a sharp rise and a decay. The peak luminosity and timescale depends on the density of the CSM and kinetic energy of the ejecta. This is natural as these observables are basically determined by diffusion of radiation inside the CSM. The fluctuation visible in the late phase of some light curves is due to the rather high Courant number taken in our numerical calculations.

\begin{figure}
\centering
\includegraphics[width=0.6\linewidth,angle=-90]{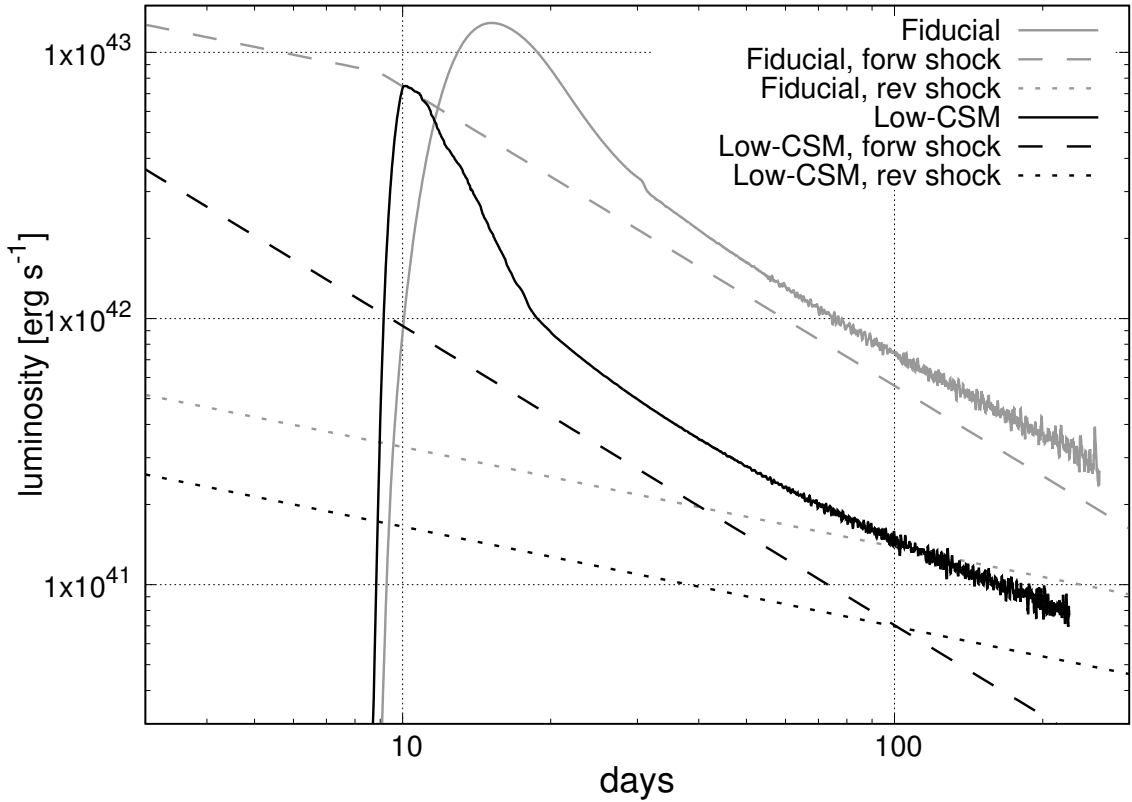}
\caption{Light curves for the first two parameter sets in Table \ref{table:IInParameters}, along with the rate of radiation released from the forward and reverse shocks, shown as dashed and dotted lines respectively. For the low-CSM case, radiation-gas equilibrium breaks down outside this figure, at $\approx 1.5$ days.}
\label{fig:lightcurves_with_shocks}
\end{figure}

To compare with the analytical model described in Section \ref{sec:semiana}, we shall see in more detail how radiation created in the forward and reverse shocks contribute to the light curve. Figure \ref{fig:lightcurves_with_shocks} shows the light curves for the fiducial and low-CSM cases, plotted with kinetic energy release at the forward and reverse shocks shown as dashed and dotted lines. To avoid the aforementioned error of the forward shock radius in the analytical model, we have used the exact radii, velocities and pressures of the two shocks numerically obtained from Chevalier's solution to draw the dashed and dotted lines. The time dependence of each component when thermal equilibrium is achieved can be calculated analytically from Chevalier's solution, and both have a dependence of $\propto t^{-3/8}$. For the forward shock, the efficiency of conversion into radiation drops as $\propto t^{-(ns-n-2s)/(n-s)}$ which is $t^{-3/4}$ for $n=10, s=2$, while the reverse shock's efficiency stays constant ($=0.6$) for the considered period. Thus at late phases, the emission from the reverse shock can become important. This is seen from Figure \ref{fig:lightcurves_with_shocks} in which the decline rate of luminosity for a model with a low mass-loss rate of $1\times10^{-2}$ M$_\odot/$yr follows the emission rate from the forward shock from day $\sim20$ to day $\sim100$ and then approaching the reverse shock component afterward.

Our numerical models also predict the early phase light curve where diffusion of radiation inside the CSM and shocked region determines the timescale and peak luminosity. The early-phase light curves predicted in this model are testable by present and future short-cadence surveys (e.g. \citealt{LSST,ASASSN,Tomoe,ZTF}).

\section{Discussion}
\label{sec:discussion}

\subsection{Dependence on Phenomenological Parameters}
\label{sec:alpha_gamma}
\begin{figure}
\centering
\includegraphics[width=0.6\linewidth,angle=-90]{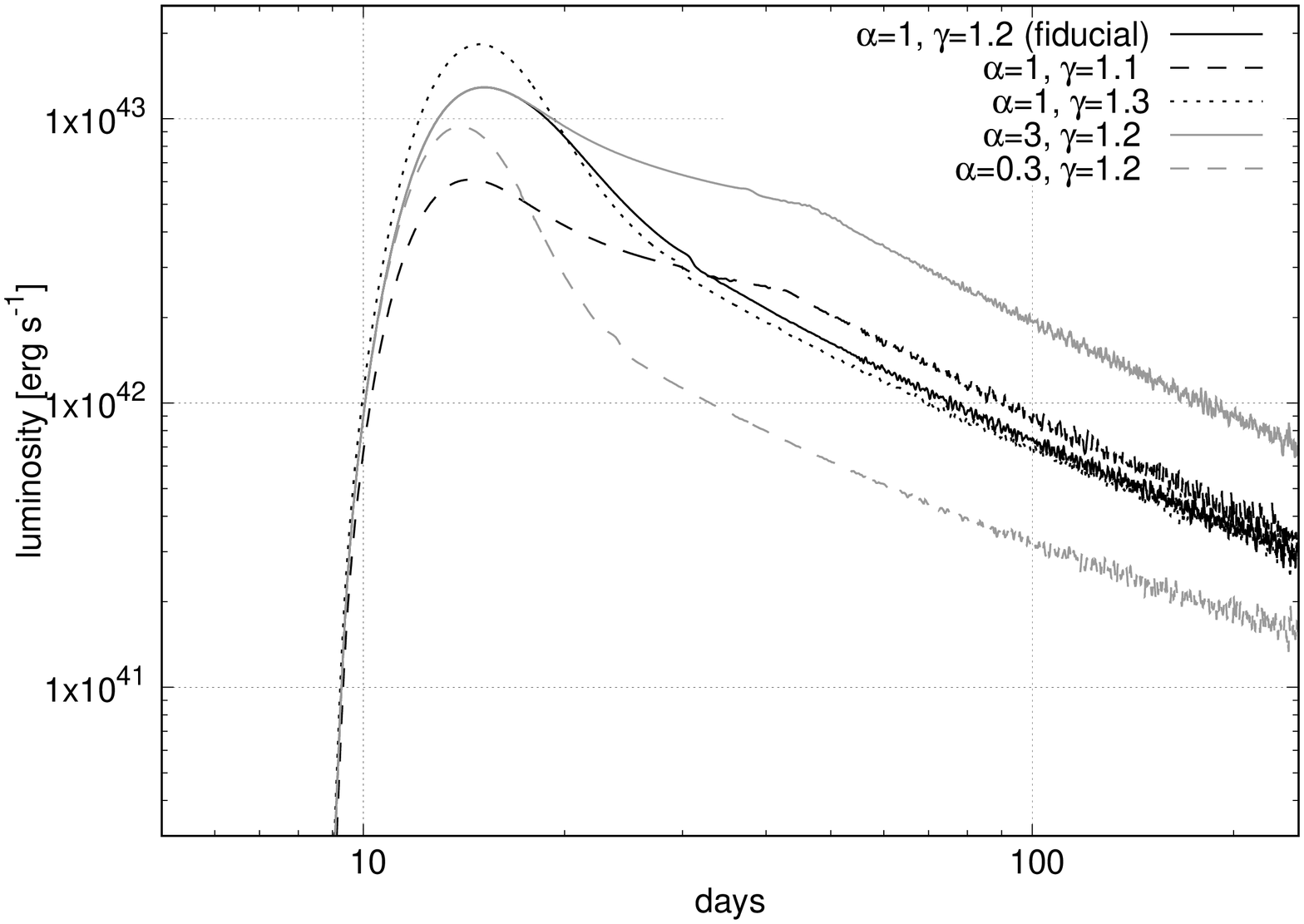}
\caption{Dependence of the light curve on the phenomenological parameters $\gamma$ and $\alpha$. We choose the parameter sets in the Fiducial case.}
\label{fig:alpha_gamma_dependence}
\end{figure}

In our model we have assumed for simplicity two constant parameters, the adiabatic index $\gamma$ and the thermalization efficiency parameter $\alpha$. Figure \ref{fig:alpha_gamma_dependence} shows the dependence of our light curve on these two parameters. 

A lower value of $\gamma$ results in a higher density in the shocked region, which results in both lower peak luminosity due to lower temperatures at the shock fronts and higher luminosity at later phases due to higher efficiency of producing radiation. Overall, we find that different values of $\gamma$ results in a factor of $\lesssim 2$ uncertainty in the peak luminosity, and much smaller uncertainties at late phases. 

A higher value of $\alpha$ results in higher luminosity at later phases when thermal equilibrium is not achieved at the forward shock. This can become a limitation when estimating the ejecta and CSM parameters from only the late phase light curve. Light curves from earlier phases around peak will be helpful to constrain this $\alpha$ parameter.

\subsection{Rayleigh-Taylor Instability}
We have claimed in this work that the density difference between the inner and outer regions may significantly affect the morphology of the light curve. However, it should be noted that this sharp density contrast can become smoothed at later phases by Rayleigh-Taylor instabilities that occur near the contact discontinuity. If we consider the interaction between the outer ejecta and CSM, the timescale of the smoothing can be crudely estimated by dimensional analysis as (e.g. \citealt{Duffell16})
\begin{eqnarray}
t_{\rm RT} &\sim& M_{\rm sh}^{5/6}E_{\rm sh}^{-1/2}\rho_{\rm CSM}^{-1/3}  \nonumber \\
& \sim & 1.3\ {\rm days} \times \left(\frac{M_{\rm sh}}{10^{-2}M_{\odot}}\right)^{5/6}\left(\frac{E_{\rm sh}}{6\times 10^{49}{\rm erg}}\right)^{-1/2}\left(\frac{\rho_{\rm CSM}}{5\times 10^{-13}{\rm g\ cm^{-3}}}\right)^{-1/3},
\end{eqnarray}
where $M_{\rm sh}, E_{\rm sh}$ are the mass and energy of the outer ejecta that is inside the shocked region respectively, and $\rho_{\rm CSM}$ is the density of the CSM at the position of the shell. For illustrative purposes we have used the values used to reproduce SN 2005ip, and the parameters are those at the start of the simulation ($t\sim0.5$ day) when the shell radius is $10^{14}$ cm. At $t\sim0.5$ day, we find that the timescale is $\sim$ 1.3 days. As these parameters evolve with time as $M_{\rm sh}\propto t^{9/10}, E_{\rm sh} \propto t^{7/10}$, and $\rho_{\rm CSM} \propto t^{-9/5}$ for $n=12, s=2$, the timescale will become longer with time in proportion to $t$, which is the same as the dynamical timescale. This same scaling with time of the two timescales is also inferred from the self-similarity in this phenomenon. 
Thus the Rayleigh-Taylor instabilities are not expected to significantly modify the shock structure for SNe like SN 2005ip.

The non-linear evolution of Rayleigh-Taylor instability was also investigated by 2D hydrodynamic simulations (\citealt{CBE92}, see also \citealt{Chevalier95}), which found instabilities mainly developing around the contact surface but not affecting the overall dynamics. Since radiation is generated in the two shocks, the instabilities will not have a large effect on the emission, especially in the optically thin regime when radiation free streams.

Nevertheless, the instability can still lead to partial smoothening of the density gradient within the shocked region. Work with multi-dimensional simulations may elaborate the effects of this on the emission.
 
\subsection{Caveats of Our Simulations}
Our numerical calculation of the light curve involves various approximations, and there are clearly various branches of improvement. 

First, our calculation assumes that LTE holds throughout the shocked region. Although this may be true in the shocked region, this is certainly not achieved in the unshocked CSM, with radiation and gas having different temperatures. Nevertheless the unshocked CSM is expected to become radiation dominated, and we expect that considering decoupling of gas and radiation will only lead to a small modification of our results. A study of the light curve relaxing the LTE approximation in the unshocked CSM is done in a different work (Takei \& Shigeyama 2019, in prep).

Second, our calculation gives a bolometric light curve but cannot follow the color evolution, especially at late times when the radiation is not thermalized. The information on color can be obtained by multi-group radiation transfer calculations, using a frequency-dependent opacity instead of the Rosseland mean. Implementing this will enable us to obtain the color and possibly the spectrum, which can realize more systematic comparisons between the theoretical model and observations of Type IIn SNe.

Third, our calculation cannot probe the regime after the inner ejecta region reaches the reverse shock. The luminosity should gradually become lower as the rate of kinetic energy released from the inner ejecta is lower, but nevertheless obtaining the light curve at this phase can enable us to probe cases where the CSM is even heavier or the ejecta is lighter than the cases we have studied here. To probe this regime we have to abandon the simple Chevalier solution, and instead numerically solve the hydrodynamical equations. Such a calculation is straightforward to implement, and we defer this to future work.

\section{Conclusions}
\label{sec:conclusion}
We have constructed a semi-analytical model of the bolometric light curve of interaction-powered supernovae. The model has two components in the emission, which are supplied by the forward and reverse shocks. The two shocks make distinct contributions, due to the forward shock more quickly becoming inefficient at converting the kinetic energy to radiation. We compared our model with the observed light curve of a well studied Type IIn supernova 2005ip. We obtain the required mass-loss rate to be roughly an order of magnitude higher than what was obtained previously in \citet{Moriya13}. This strengthens the importance of considering the two components to accurately understand the mass-loss history of the progenitor.

Although our analytical model allows us to relatively easily extract the parameters of the ejecta and CSM from observations, it is only applicable to the later phases of the light curve when diffusion time inside the CSM becomes negligible. Our numerical calculations are able to predict the early rise of the light curve, that comes from the diffusion of radiation in the CSM. For testing our models the information of the light curve from both early and late phases will be important. This is feasible with present and future short-cadence surveys.

The ultimate goal in this one-dimensional scheme would be to conduct radiation hydrodynamics simulations with sufficient resolution in the shocked region. In Section \ref{sec:discussion} we have seen that our results do not depend sensitively on the adiabatic index, if this is set to constant as a function of time and radius. Nevertheless, the strength of radiation feedback changing over time and radius may leave important imprints on the light curve. Understanding this would require much more computational cost, and is deferred to future work.

\acknowledgements
The authors thank the anonymous referee for giving important comments on the manuscript. The authors also thank Takashi J. Moriya for providing us the bolometric light curve data for SN 2005ip, and giving us valuable comments on this manuscript. DT thanks Yuki Takei for fruitful discussions over afternoon coffee. DT is supported by the Advanced Leading Graduate Course for Photon Science (ALPS) at the University of Tokyo.
This work is also supported by JSPS KAKENHI Grant Numbers JP19J21578, JP17K14248, JP18H04573, 16H06341, 16K05287, 15H02082, MEXT, Japan.

\bibliography{IIn}{}

\begin{thebibliography}{44}
\expandafter\ifx\csname natexlab\endcsname\relax\def\natexlab#1{#1}\fi

\bibitem[{{Aretxaga} {et~al.}(1999){Aretxaga}, {Benetti}, {Terlevich},
  {Fabian}, {Cappellaro}, {Turatto}, \& {della Valle}}]{Aretxaga99}
{Aretxaga}, I., {Benetti}, S., {Terlevich}, R.~J., {Fabian}, A.~C.,
  {Cappellaro}, E., {Turatto}, M., \& {della Valle}, M. 1999, \mnras, 309, 343

\bibitem[{{Chatzopoulos} {et~al.}(2012){Chatzopoulos}, {Wheeler}, \&
  {Vinko}}]{Chatzopoulos12}
{Chatzopoulos}, E., {Wheeler}, J.~C., \& {Vinko}, J. 2012, \apj, 746, 121

\bibitem[{{Chevalier} \& {Blondin}(1995)}]{Chevalier95}
{Chevalier}, R. \& {Blondin}, J.~M. 1995, \apj, 444, 312

\bibitem[{{Chevalier}(1982)}]{Chevalier82}
{Chevalier}, R.~A. 1982, \apj, 258, 790

\bibitem[{{Chevalier}(2012)}]{Chevalier12}
---. 2012, \apjl, 752, L2

\bibitem[{{Chevalier} {et~al.}(1992){Chevalier}, {Blondin}, \&
  {Emmering}}]{CBE92}
{Chevalier}, R.~A., {Blondin}, J.~M., \& {Emmering}, R.~T. 1992, \apj, 392, 118

\bibitem[{{Chevalier} \& {Fransson}(2003)}]{Chevalier03}
{Chevalier}, R.~A. \& {Fransson}, C. 2003, in Lecture Notes in Physics, Berlin
  Springer Verlag, Vol. 598, Supernovae and Gamma-Ray Bursters, ed.
  K.~{Weiler}, 171--194

\bibitem[{{Chevalier} \& {Soker}(1989)}]{CS89}
{Chevalier}, R.~A. \& {Soker}, N. 1989, \apj, 341, 867

\bibitem[{{Chugai}(1991)}]{Chugai91}
{Chugai}, N.~N. 1991, \mnras, 250, 513

\bibitem[{{Chugai} \& {Danziger}(1994)}]{Chugai94}
{Chugai}, N.~N. \& {Danziger}, I.~J. 1994, \mnras, 268, 173

\bibitem[{{Dessart} {et~al.}(2015){Dessart}, {Audit}, \& {Hillier}}]{Dessart15}
{Dessart}, L., {Audit}, E., \& {Hillier}, D.~J. 2015, \mnras, 449, 4304

\bibitem[{{Duffell}(2016)}]{Duffell16}
{Duffell}, P.~C. 2016, \apj, 821, 76

\bibitem[{{Elias-Rosa} {et~al.}(2018){Elias-Rosa}, {Van Dyk}, {Benetti},
  {Cappellaro}, {Smith}, {Kotak}, {Turatto}, {Filippenko}, {Pignata}, {Fox},
  {Galbany}, {Gonz{\'a}lez-Gait{\'a}n}, {Miluzio}, {Monard}, \&
  {Ergon}}]{Elias-Rosa18}
{Elias-Rosa}, N., {Van Dyk}, S.~D., {Benetti}, S., {Cappellaro}, E., {Smith},
  N., {Kotak}, R., {Turatto}, M., {Filippenko}, A.~V., {Pignata}, G., {Fox},
  O.~D., {Galbany}, L., {Gonz{\'a}lez-Gait{\'a}n}, S., {Miluzio}, M., {Monard},
  L.~A.~G., \& {Ergon}, M. 2018, \apj, 860, 68

\bibitem[{{Filippenko}(1997)}]{Filippenko97}
{Filippenko}, A.~V. 1997, Annual Review of Astronomy and Astrophysics, 35, 309

\bibitem[{{Fox} {et~al.}(2010){Fox}, {Chevalier}, {Dwek}, {Skrutskie},
  {Sugerman}, \& {Leisenring}}]{Fox10}
{Fox}, O.~D., {Chevalier}, R.~A., {Dwek}, E., {Skrutskie}, M.~F., {Sugerman},
  B. E.~K., \& {Leisenring}, J.~M. 2010, \apj, 725, 1768

\bibitem[{{Fox} {et~al.}(2011){Fox}, {Chevalier}, {Skrutskie}, {Soderberg},
  {Filippenko}, {Ganeshalingam}, {Silverman}, {Smith}, \& {Steele}}]{Fox11}
{Fox}, O.~D., {Chevalier}, R.~A., {Skrutskie}, M.~F., {Soderberg}, A.~M.,
  {Filippenko}, A.~V., {Ganeshalingam}, M., {Silverman}, J.~M., {Smith}, N., \&
  {Steele}, T.~N. 2011, \apj, 741, 7

\bibitem[{{Graham} {et~al.}(2019){Graham}, {Kulkarni}, {Bellm}, {Adams},
  {Barbarino}, {Blagorodnova}, {Bodewits}, {Bolin}, {Brady}, \& {Cenko}}]{ZTF}
{Graham}, M.~J., {Kulkarni}, S.~R., {Bellm}, E.~C., {Adams}, S.~M.,
  {Barbarino}, C., {Blagorodnova}, N., {Bodewits}, D., {Bolin}, B., {Brady},
  P.~R., \& {Cenko}, S.~B. 2019, \pasp, 131, 078001

\bibitem[{{Grasberg} \& {Nadezhin}(1986)}]{Grasberg86}
{Grasberg}, E.~K. \& {Nadezhin}, D.~K. 1986, Pisma v Astronomicheskii Zhurnal,
  12, 168

\bibitem[{{Iglesias} \& {Rogers}(1996)}]{Iglesias96}
{Iglesias}, C.~A. \& {Rogers}, F.~J. 1996, \apj, 464, 943

\bibitem[{{Katsuda} {et~al.}(2014){Katsuda}, {Maeda}, {Nozawa}, {Pooley}, \&
  {Immler}}]{Katsuda14}
{Katsuda}, S., {Maeda}, K., {Nozawa}, T., {Pooley}, D., \& {Immler}, S. 2014,
  \apj, 780, 184

\bibitem[{{Kiewe} {et~al.}(2012){Kiewe}, {Gal-Yam}, {Arcavi}, {Leonard},
  {Emilio Enriquez}, {Cenko}, {Fox}, {Moon}, {Sand }, {Soderberg}, \&
  {CCCP}}]{Kiewe12}
{Kiewe}, M., {Gal-Yam}, A., {Arcavi}, I., {Leonard}, D.~C., {Emilio Enriquez},
  J., {Cenko}, S.~B., {Fox}, D.~B., {Moon}, D.-S., {Sand }, D.~J., {Soderberg},
  A.~M., \& {CCCP}, T. 2012, \apj, 744, 10

\bibitem[{{Levermore} \& {Pomraning}(1981)}]{LP81}
{Levermore}, C.~D. \& {Pomraning}, G.~C. 1981, \apj, 248, 321

\bibitem[{{LSST Science Collaboration} {et~al.}(2009){LSST Science
  Collaboration}, {Abell}, {Allison}, {Anderson}, {Andrew}, {Angel}, {Armus},
  {Arnett}, {Asztalos}, \& {Axelrod}}]{LSST}
{LSST Science Collaboration}, {Abell}, P.~A., {Allison}, J., {Anderson}, S.~F.,
  {Andrew}, J.~R., {Angel}, J. R.~P., {Armus}, L., {Arnett}, D., {Asztalos},
  S.~J., \& {Axelrod}, T.~S. 2009, arXiv e-prints, arXiv:0912.0201

\bibitem[{{Matzner} \& {McKee}(1999)}]{MM99}
{Matzner}, C.~D. \& {McKee}, C.~F. 1999, \apj, 510, 379

\bibitem[{{Moriya} {et~al.}(2011){Moriya}, {Tominaga}, {Blinnikov}, {Baklanov},
  \& {Sorokina}}]{Moriya11}
{Moriya}, T., {Tominaga}, N., {Blinnikov}, S.~I., {Baklanov}, P.~V., \&
  {Sorokina}, E.~I. 2011, \mnras, 415, 199

\bibitem[{{Moriya}(2014)}]{Moriya14b}
{Moriya}, T.~J. 2014, \aap, 564, A83

\bibitem[{{Moriya} {et~al.}(2013){Moriya}, {Maeda}, {Taddia}, {Sollerman},
  {Blinnikov}, \& {Sorokina}}]{Moriya13}
{Moriya}, T.~J., {Maeda}, K., {Taddia}, F., {Sollerman}, J., {Blinnikov},
  S.~I., \& {Sorokina}, E.~I. 2013, \mnras, 435, 1520

\bibitem[{{Moriya} {et~al.}(2014){Moriya}, {Maeda}, {Taddia}, {Sollerman},
  {Blinnikov}, \& {Sorokina}}]{Moriya14a}
---. 2014, \mnras, 439, 2917

\bibitem[{{Nakar} \& {Sari}(2010)}]{Nakar10}
{Nakar}, E. \& {Sari}, R. 2010, \apj, 725, 904

\bibitem[{{Quataert} \& {Shiode}(2012)}]{QS12}
{Quataert}, E. \& {Shiode}, J. 2012, \mnras, 423, L92

\bibitem[{{Rybicki} \& {Lightman}(1979)}]{Radipro}
{Rybicki}, G.~B. \& {Lightman}, A.~P. 1979, {Radiative processes in
  astrophysics} (Wiley-VCH)

\bibitem[{{Sako} {et~al.}(2016){Sako}, {Osawa}, {Takahashi}, {Kikuchi}, {Doi},
  {Kobayashi}, {Aoki}, {Arimatsu}, {Ichiki}, {Ikeda}, {Ita}, {Kasuga},
  {Kawakita}, {Kokubo}, {Maehara}, {Matsunaga}, {Mito}, {Mitsuda}, {Miyata},
  {Mori}, {Mori}, {Morii}, {Morokuma}, {Motohara}, {Nakada}, {Osawa},
  {Okumura}, {Onozato}, {Sarugaku}, {Sato}, {Shigeyama}, {Soyano}, {Tanaka},
  {Taniguchi}, {Tanikawa}, {Tarusawa}, {Tominaga}, {Totani}, {Urakawa}, {Usui},
  {Watanabe}, {Yamaguchi}, \& {Yoshikawa}}]{Tomoe}
{Sako}, S., {Osawa}, R., {Takahashi}, H., {Kikuchi}, Y., {Doi}, M.,
  {Kobayashi}, N., {Aoki}, T., {Arimatsu}, K., {Ichiki}, M., {Ikeda}, S.,
  {Ita}, Y., {Kasuga}, T., {Kawakita}, H., {Kokubo}, M., {Maehara}, H.,
  {Matsunaga}, N., {Mito}, H., {Mitsuda}, K., {Miyata}, T., {Mori}, K., {Mori},
  Y., {Morii}, M., {Morokuma}, T., {Motohara}, K., {Nakada}, Y., {Osawa}, K.,
  {Okumura}, S.-i., {Onozato}, H., {Sarugaku}, Y., {Sato}, M., {Shigeyama}, T.,
  {Soyano}, T., {Tanaka}, M., {Taniguchi}, Y., {Tanikawa}, A., {Tarusawa}, K.,
  {Tominaga}, N., {Totani}, T., {Urakawa}, S., {Usui}, F., {Watanabe}, J.,
  {Yamaguchi}, J., \& {Yoshikawa}, M. 2016, in \procspie, Vol. 9908,
  Ground-based and Airborne Instrumentation for Astronomy VI, 99083P

\bibitem[{{Salamanca} {et~al.}(1998){Salamanca}, {Cid-Fernandes},
  {Tenorio-Tagle}, {Telles}, {Terlevich}, \& {Munoz-Tunon}}]{Salamanca98}
{Salamanca}, I., {Cid-Fernandes}, R., {Tenorio-Tagle}, G., {Telles}, E.,
  {Terlevich}, R.~J., \& {Munoz-Tunon}, C. 1998, \mnras, 300, L17

\bibitem[{{Schlegel}(1990)}]{Schlegel90}
{Schlegel}, E.~M. 1990, \mnras, 244, 269

\bibitem[{{Shappee} {et~al.}(2014){Shappee}, {Prieto}, {Grupe}, {Kochanek},
  {Stanek}, {De Rosa}, {Mathur}, {Zu}, {Peterson}, \& {Pogge}}]{ASASSN}
{Shappee}, B.~J., {Prieto}, J.~L., {Grupe}, D., {Kochanek}, C.~S., {Stanek},
  K.~Z., {De Rosa}, G., {Mathur}, S., {Zu}, Y., {Peterson}, B.~M., \& {Pogge},
  R.~W. 2014, \apj, 788, 48

\bibitem[{{Shiode} \& {Quataert}(2014)}]{SQ14}
{Shiode}, J.~H. \& {Quataert}, E. 2014, \apj, 780, 96

\bibitem[{{Smith} {et~al.}(2011){Smith}, {Li}, {Filippenko}, \&
  {Chornock}}]{Smith11}
{Smith}, N., {Li}, W., {Filippenko}, A.~V., \& {Chornock}, R. 2011, \mnras,
  412, 1522

\bibitem[{{Smith} {et~al.}(2009){Smith}, {Silverman}, {Chornock}, {Filippenko},
  {Wang}, {Li}, {Ganeshalingam}, {Foley}, {Rex}, \& {Steele}}]{Smith09}
{Smith}, N., {Silverman}, J.~M., {Chornock}, R., {Filippenko}, A.~V., {Wang},
  X., {Li}, W., {Ganeshalingam}, M., {Foley}, R.~J., {Rex}, J., \& {Steele},
  T.~N. 2009, \apj, 695, 1334

\bibitem[{{Soumagnac} {et~al.}(2019){Soumagnac}, {Ofek}, {Gal-yam}, {Waxman},
  {Ginzburg}, {Linn Strotjohann}, {Schulze}, {Barlow}, {Behar}, {Chelouche},
  {Fremling}, {Ganot}, {Gezari}, {Kasliwal}, {Kaspi}, {Kulkarni}, {Laher},
  {Maoz}, {Martin}, {Nakar}, {Neill}, {Nugent}, {Poznanski}, \&
  {Yaron}}]{Soumagnac19}
{Soumagnac}, M.~T., {Ofek}, E.~O., {Gal-yam}, A., {Waxman}, E., {Ginzburg}, S.,
  {Linn Strotjohann}, N., {Schulze}, S., {Barlow}, T.~A., {Behar}, E.,
  {Chelouche}, D., {Fremling}, C., {Ganot}, N., {Gezari}, S., {Kasliwal},
  M.~M., {Kaspi}, S., {Kulkarni}, S.~R., {Laher}, R.~R., {Maoz}, D., {Martin},
  C.~D., {Nakar}, E., {Neill}, J.~D., {Nugent}, P.~E., {Poznanski}, D., \&
  {Yaron}, O. 2019, \apj, 872, 141

\bibitem[{{Stritzinger} {et~al.}(2012){Stritzinger}, {Taddia}, {Fransson},
  {Fox}, {Morrell}, {Phillips}, {Sollerman}, {Anderson}, {Boldt}, {Brown},
  {Campillay}, {Castellon}, {Contreras}, {Folatelli}, {Habergham}, {Hamuy},
  {Hjorth}, {James}, {Krzeminski}, {Mattila}, {Persson}, \&
  {Roth}}]{Stritzinger12}
{Stritzinger}, M., {Taddia}, F., {Fransson}, C., {Fox}, O.~D., {Morrell}, N.,
  {Phillips}, M.~M., {Sollerman}, J., {Anderson}, J.~P., {Boldt}, L., {Brown},
  P.~J., {Campillay}, A., {Castellon}, S., {Contreras}, C., {Folatelli}, G.,
  {Habergham}, S.~M., {Hamuy}, M., {Hjorth}, J., {James}, P.~A., {Krzeminski},
  W., {Mattila}, S., {Persson}, S.~E., \& {Roth}, M. 2012, \apj, 756, 173

\bibitem[{{Taddia} {et~al.}(2013){Taddia}, {Stritzinger}, {Sollerman},
  {Phillips}, {Anderson}, {Boldt}, {Campillay}, {Castell{\'o}n}, {Contreras},
  {Folatelli}, {Hamuy}, {Heinrich-Josties}, {Krzeminski}, {Morrell}, {Burns},
  {Freedman}, {Madore}, {Persson}, \& {Suntzeff}}]{Taddia13}
{Taddia}, F., {Stritzinger}, M.~D., {Sollerman}, J., {Phillips}, M.~M.,
  {Anderson}, J.~P., {Boldt}, L., {Campillay}, A., {Castell{\'o}n}, S.,
  {Contreras}, C., {Folatelli}, G., {Hamuy}, M., {Heinrich-Josties}, E.,
  {Krzeminski}, W., {Morrell}, N., {Burns}, C.~R., {Freedman}, W.~L., {Madore},
  B.~F., {Persson}, S.~E., \& {Suntzeff}, N.~B. 2013, \aap, 555, A10

\bibitem[{{van Marle} {et~al.}(2010){van Marle}, {Smith}, {Owocki}, \& {van
  Veelen}}]{vanMarle10}
{van Marle}, A.~J., {Smith}, N., {Owocki}, S.~P., \& {van Veelen}, B. 2010,
  \mnras, 407, 2305

\bibitem[{{Vlasis} {et~al.}(2016){Vlasis}, {Dessart}, \& {Audit}}]{Vlasis16}
{Vlasis}, A., {Dessart}, L., \& {Audit}, E. 2016, \mnras, 458, 1253

\bibitem[{{Woosley} {et~al.}(2007){Woosley}, {Blinnikov}, \& {Heger}}]{WBH07}
{Woosley}, S.~E., {Blinnikov}, S., \& {Heger}, A. 2007, \nat, 450, 390

\end{thebibliography}
\bibliographystyle{apj}



\end{document}